\documentclass[referee]{aa} 
\input psfig.sty
\begin{document}
\title{Radio observations and spectral index study of SNR G126.2+1.6}
\author{Wenwu Tian \inst{1,2}
\and
      Denis Leahy \inst{2}}
\authorrunning{W.W. Tian and D.A. Leahy}
\offprints{W. W. Tian}
\institute{National Astronomical Observatories, CAS, Beijing 100012, China \\
\and
Department of Physics \& Astronomy, University of Calgary, Calgary, Alberta T2N 1N4, Canada}
 
\date{Received Aug. 11, 2005; accepted XX, 2005} 
\abstract{
We present new images of the low radio surface brightness Supernova Remnant (SNR) G126.2+1.6, based on the 408 MHz and 1420 MHz continuum emission and the HI-line emission data of the Canadian Galactic Plane Survey (CGPS). 
 We find the SNR's flux densities at 408 MHz (9.7$\pm$3.9 Jy) and 1420 MHz (6.7$\pm$2.1 Jy) which have been corrected for flux densities from compact sources within the SNR.  
The integrated flux density based spectral index (S$_{\nu}$$\propto$$\nu$$^{-\alpha}$)  is  0.30$\pm$0.41.  The respective T-T plot 
spectral index is 0.30 $\pm$0.08.
We also find spatial variations of spectral index within the SNR:~0.2 - ~0.6. HI observations show structures probably associated with the SNR, i.e,
features associated with the SNR's southeastern filaments in the radial velocity range of -33 to -42 km$/$s, and with its northwestern filaments in -47 to -52 km$/$s. This association suggests a distance of 5.6 kpc for SNR G126.2+1.6.  The estimated Sedov age for G126.2+1.6 is less than 2.1$\times$10$^{5}$ yr.

\keywords{ISM:individual (G126.2+1.6) - radio continuum - HI-line:ISM}}
\titlerunning{Radio Spectrum of the SNR G126.2+1.6}
\maketitle 

\section{Introduction}
As the third of a continuing study of supernova remnants 
spectral index variation (Tian $\&$ Leahy 2005, Leahy $\&$ Tian 2005), the radio  spectrum 
of the SNR G126.2+1.6 is studied in detail in the paper. This low radio surface brightness 
source was first detected and classified as a SNR by Reich et al. (1979). Some of its  basic 
physical features are unclear, such as its distance and radio spectrum. Flux densities 
of G126.2+1.6 and respective spectral indices have been given previously (Reich et al. 
2003; Joncas et al. 1989, F\"urst et al. 1984) based on observations with a 
lower sensitivity than the current observations. Previous studies show that better flux 
density measurements are needed for G126.2+1.6. In the paper, we present the SNR's 
continuum images at higher sensitivity than previously at 408 MHz and 1420 MHz in 
order to determine its flux densities and spectral index. We investigate HI-line 
emission at various radial velocities for detecting interactions of the remnant 
with the surrounding gas and estimating its distance and age. 

\section{Observations and Analysis}

The continuum and HI emission data sets come from the CGPS,
which is described in detail by Taylor et al. (2003).
The data sets are mainly based on observations from the Synthesis Telescope 
(ST) of the Dominion Radio Astrophysical Observatory (DRAO). The spatial
resolution of the continuum images is better than 1$^{\prime}$$\times$ 1$^{\prime}$ cosec($\delta$) (HPBW) 
at 1420 MHz and 3.4$^{\prime}$$\times$3.4$^{\prime}$ cosec($\delta$) at 408 MHz. The synthesized beam for 
the HI line images is as the same as for the continuum  and the radial velocity 
resolution is 1.32 km$/$s. DRAO ST observations 
are not sensitive to structures larger than an angular 
size scale of about 3.3$^{o}$ at 408 MHz and 56$^{\prime}$  at 1420 MHz. Thus the CGPS includes 
data from the 408 MHz all-sky survey of Haslam et al (1982), sensitive to structure greater 51$^{\prime}$, and the Effelsberg 1.4 GHz Galactic plane survey 
of Reich et al. (1990, 1997), sensitive to structure with 9.4$^{\prime}$ for large scale emission 
(the single-dish data are freely available by http://www.mpifr-bonn.mpg.de/survey.html). 
The low-order spacing HI data is from the single-antenna survey of the 
CGPS area (Higgs $\&$ Tapping 2000) with resolution of 36$^{\prime}$.  See Taylor et al. (2003) for detail of the method of combining the synthesis telescopes and single dish observations. 

We analyze the continuum and HI images of G126.2+1.6 and determine its flux
densities and distance using the DRAO export software package.  
 For G126.2+1.6, integrated flux density's errors are found by comparing results for several different choices of background region. For compact sources, the flux densitiy's errors are taken as the formal Gaussian fit errors.  
The influence of compact sources within the SNR is  
reduced by employing similar methods to Tian and Leahy (2005).
 
\section{Results}
\begin{figure*}
\vspace{55mm}
\begin{picture}(200,300)
\put(-30,145){\includegraphics{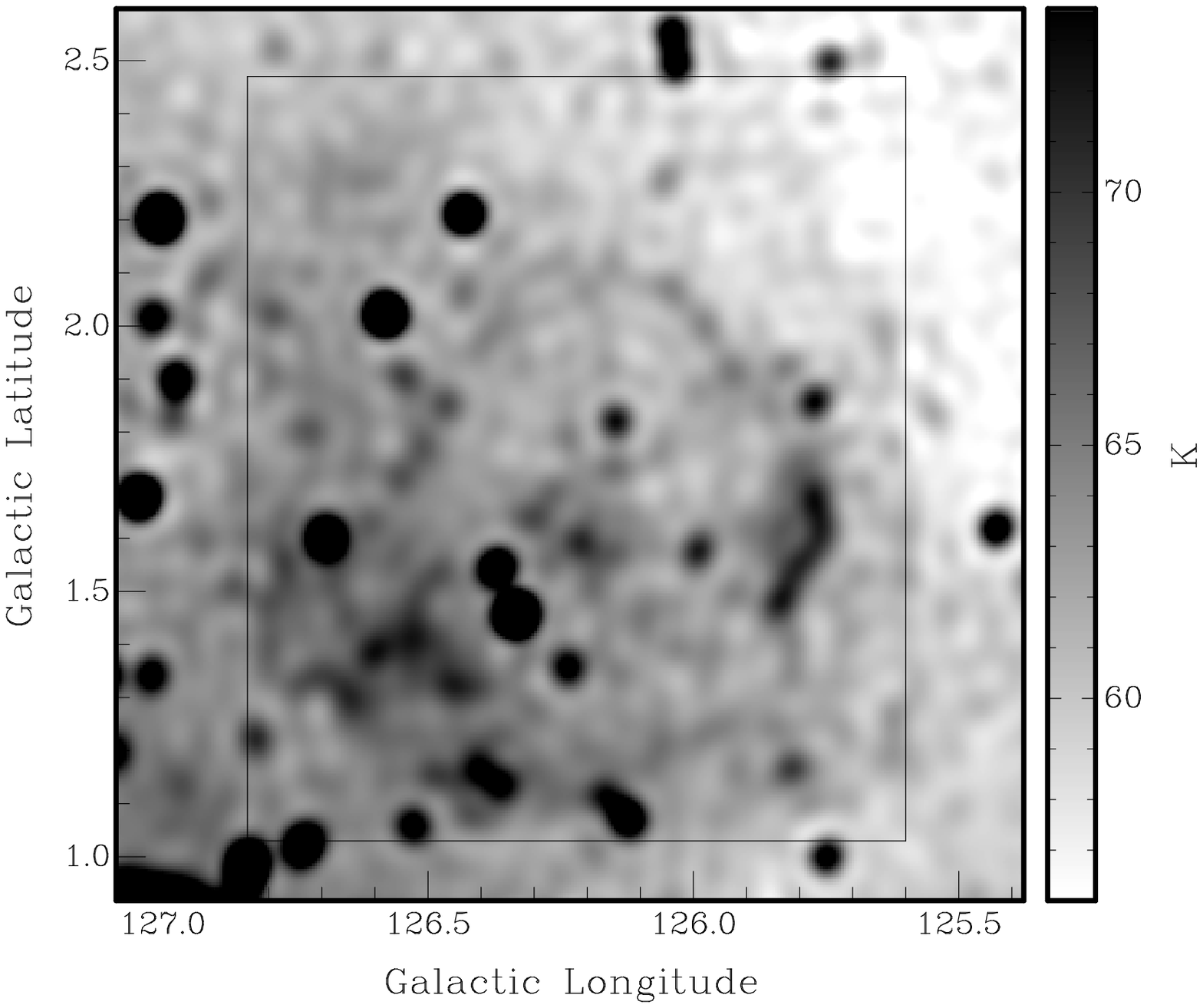}}
\put(225,145){\includegraphics{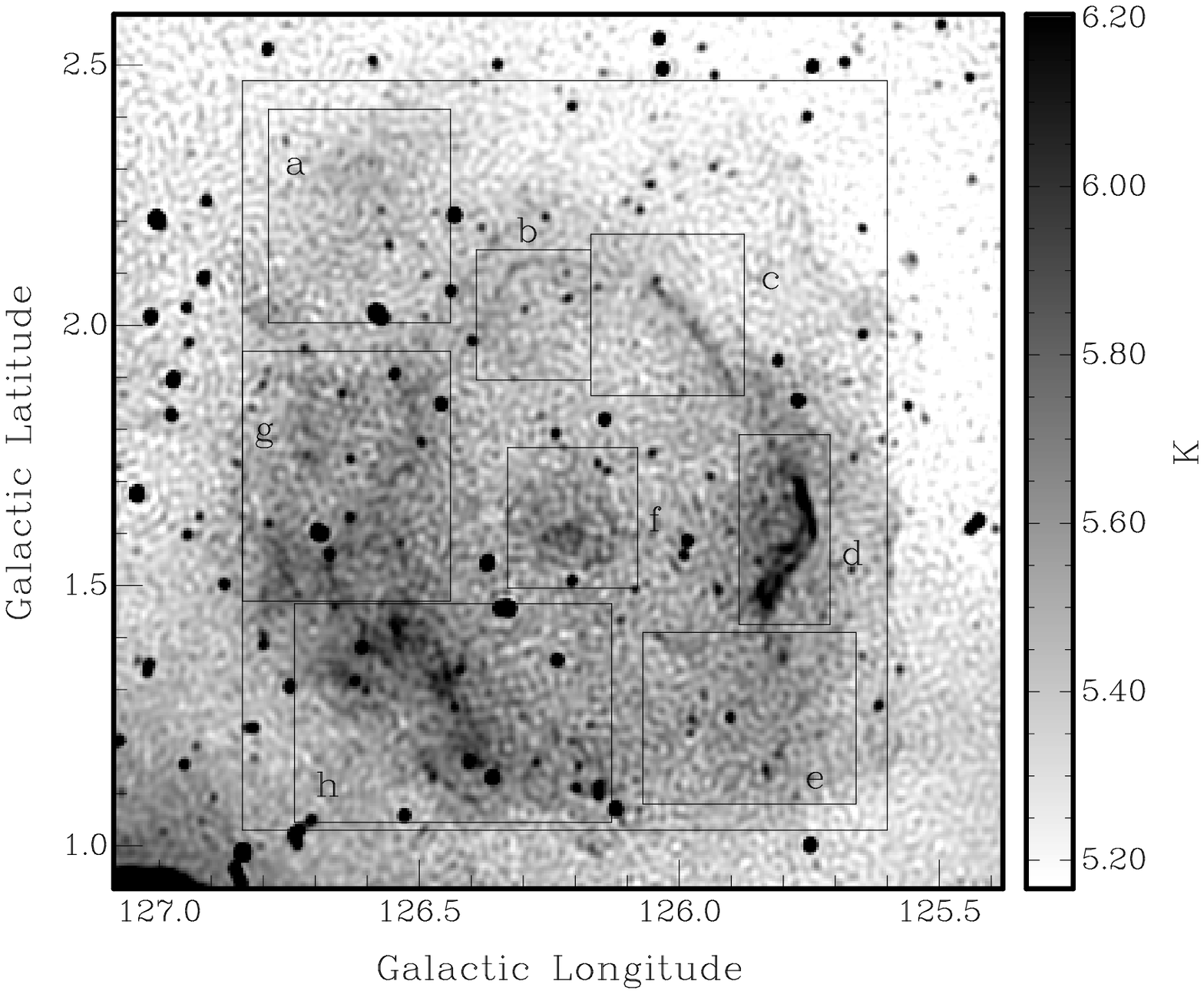}}
\put(-100,-185){\includegraphics{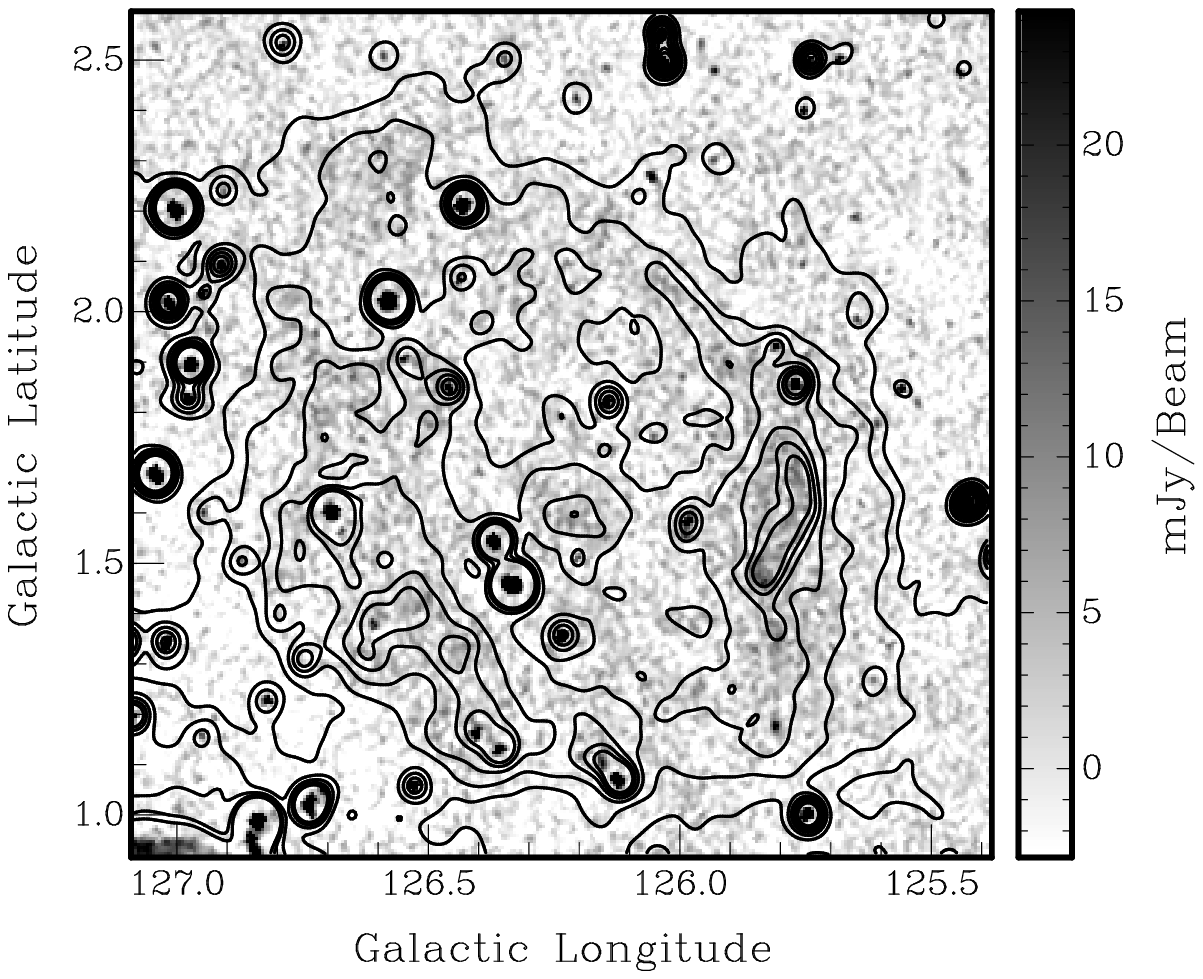}}
\put(211,-111){\includegraphics{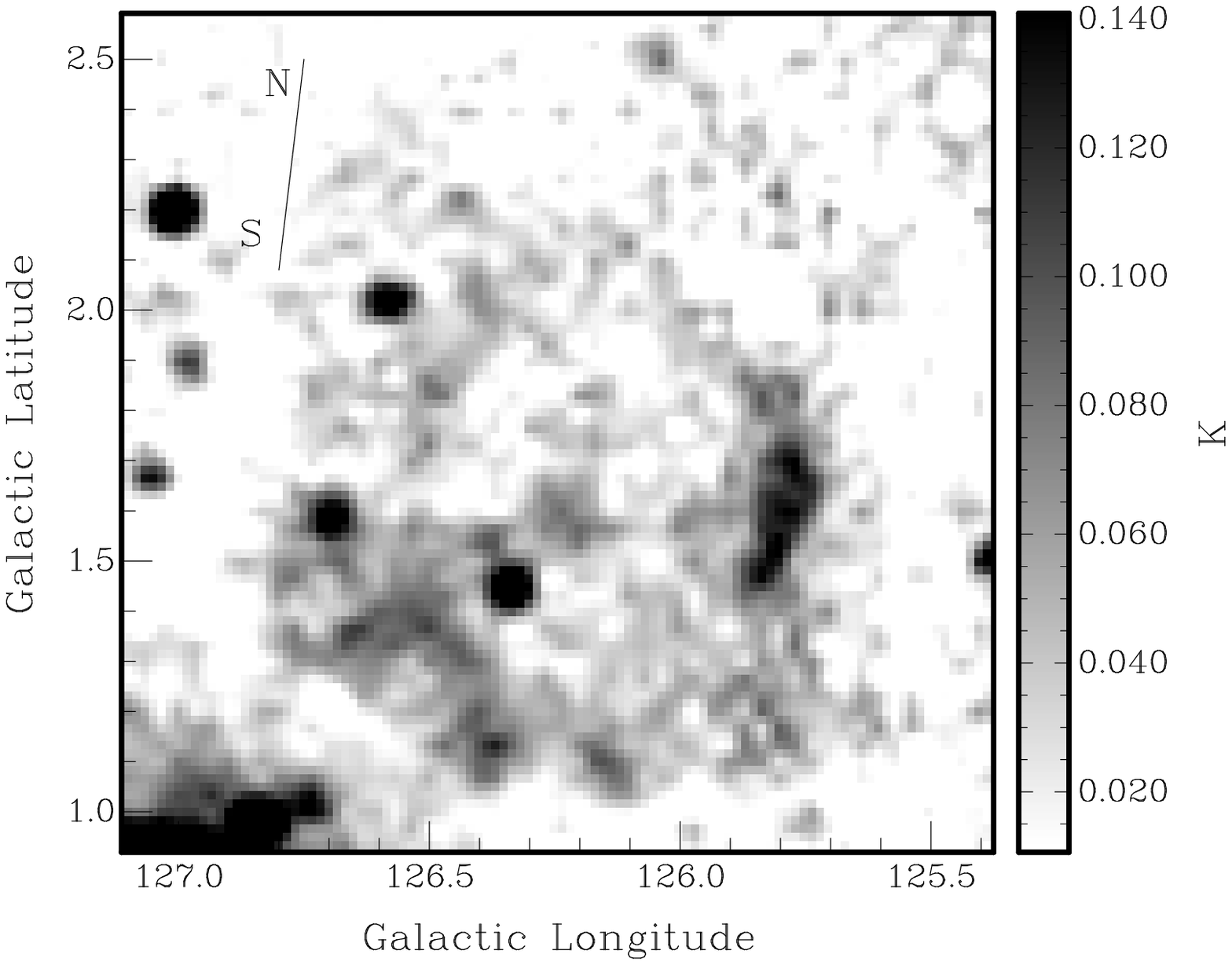}}
\end{picture}
\caption[xx]{The first row of images shows the CGPS maps at 
408 MHz (left) and 1420 MHz (right). The left of the second row shows the WENSS image at 
327 MHz (grey scale) with contours from the 1420 MHz map convolved to the same resolution 
as the 408 MHz resolution (contours at 5.3, 5.4, 5.5, 5.6, 5.7, 5.8, 5.9 mJy$/$beam). 
The right of the second row is the 
2695 MHz Effelsberg map. The box used for whole SNR T-T plots 
is shown in the upper left. The 8 boxes, labeled with letters and 
used for SNR sub-areas T-T plots, are shown in the upper right.
The direction of North (N) and South (S) is marked on the lower right image.}
\end{figure*}

\subsection{Continuum Emission from G126.2+1.6}

The CGPS continuum images at 408 MHz and 1420 MHz are shown in the upper left and right
panels of Fig. 1. 
The lower left shows the WENSS (the Westerbork North Sky Survey, beam size  
54 $^{\prime \prime}$$\times $ 54 $^{\prime \prime}$ cosec($\delta$),  Rengelink R.B. et al. 1997) map at 327 MHz 
(grey scale) with contours from the 1420 MHz map convolved to the same resolution 
as the 408 MHz map. 
The 2695 MHz Effelsberg map is reproduced in the lower right for reference
(F\"urst et al., 1990). The Effelsberg map has an resolution of 4.3$^{\prime}$ and a sensitivity of 
50 mKT$_{B}$.

The 408 MHz image for G126.2+1.6 reveals clearly more structure in comparison with 
Joncas's (1989) 408 MHz image, 
including a ring-like filamentary structure. 
Sharp and diffuse filamentary emissions are detected around the limb of G126.2+1.6, especially the bright western and northwestern 
sharp filaments and south to southeastern diffuse filamentary emissions.
An area located at the center of the SNR shows significant diffuse emission.   
Compact sources distributed across the face of the SNR area are prominent, and 
illustrate the importance of having high enough spatial resolution to distinguish
between SNR and compact source emission.
The 1420 MHz map is the first published at this frequency and the best so far at any 
frequencies for G126.2+1.6. It shows much better detail of the fine structure in 
G126.2+1.6 than any previous images. 
A double filament structure appears in the northeastern part of the SNR, and 
multi-filamentary 
structures appear in the southeastern part. The inside edge of the filament distribution 
has a diamond-type shape more than a ring-type shape. 
The outlines of G126.2+1.6 at 408 MHz and 1420 MHz are very similar. The lower left of 
Fig. 1 shows that main features of G126.2+1.6 in the 327 MHz WENSS map are consistent 
with the CGPS 408 MHz and 1420 MHz maps. 

\begin{table*}
\begin{center}
\caption{List of brightest compact sources and their integrated Flux Densities (FD) inside G126.2+1.6} \setlength{\tabcolsep}{1mm} \begin{tabular}{cccccccc} \hline \hline
 Source Num.&RA(2000) & Dec(2000) &GLONG& GLAT & FD at 408MHz& FD at 1420MHz& Sp. Index\\ 
\hline & [h m s]&[$^{\circ}$ ' "]&deg &deg &mJy&mJy&$\alpha$\\ 
\hline 
\hline 
1& 1 22 39.55&64  7 24.2&  126.331&1.458 &412 $\pm$15&153$\pm$ 5&0.80 ( 0.76 to 0.84 )\\ 
2& 1 25 35.93&64 39 16.0&  126.579&2.025 &336 $\pm$11&132$\pm$ 4&0.75 ( 0.72 to 0.79 )\\ 
3& 1 24 26.25&64 51 48.0&  126.429&2.216 &223 $\pm$ 7& 63$\pm$ 2&1.02 ( 0.98 to 1.05 )\\
4& 1 26  5.20&64 13 14.1&  126.689&1.602 &218 $\pm$ 7& 91$\pm$ 3&0.70 ( 0.66 to 0.74 )\\
5& 1 23  5.06&64 12 19.3&  126.367&1.545 &137 $\pm$9 & 54$\pm$ 3&0.75 ( 0.69 to 0.82 )\\
6& 1 22 48.99&63 48 39.6&  126.386&1.150 &134 $\pm$ 9& 82$\pm$11&0.39 ( 0.28 to 0.52 )\\
7& 1 20 19.86&63 45 45.7&  126.119&1.070 &121 $\pm$ 5& 38$\pm$ 2&0.93 ( 0.88 to 0.98 )\\
8& 1 23 58.46&63 41 53.8&  126.527&1.054 & 63 $\pm$ 4& 22$\pm$ 1&0.86 ( 0.80 to 0.94 )\\
9& 1 21 38.46&64  2  6.0&  126.231&1.357 & 63 $\pm$ 4& 16$\pm$ 1&1.08 ( 1.00 to 1.16 )\\
10&1 20 40.53&63 47 47.8&  126.153&1.108 & 58 $\pm$ 5& 26$\pm$ 2&0.65 ( 0.57 to 0.74 )\\
11&1 17 53.45&64 34 47.7&  125.767&1.854 & 54 $\pm$ 5& 32$\pm$ 3&0.43 ( 0.33 to 0.54 )\\
12&1 19 37.81&64 17  3.8&  125.985&1.580 & 49 $\pm$ 5& 19$\pm$ 2&0.75 ( 0.64 to 0.86 )\\
13&1 21 19.04&64 30 29.9&  126.141&1.823 & 46 $\pm$ 4& 20$\pm$ 1&0.66 ( 0.59 to 0.74 )\\
14&1 22 31.42&63 47 37.6&  126.356&1.129 & 37 $\pm$ 8& 16$\pm$ 3&0.65 ( 0.46 to 0.92 )\\
15&1 26 49.86&63 49 40.5&  126.823&1.224 & 37 $\pm$ 3& 15$\pm$ 3&0.74 ( 0.59 to 0.92 )\\
16&1 24 31.14&64  2 54.6&  126.542&1.409 & 33 $\pm$ 5& 12$\pm$ 3&0.79 ( 0.58 to 1.08 )\\
17&1 24 15.65&64 29 48.3&  126.457&1.850 & 28 $\pm$ 6& 21$\pm$ 2&0.22 ( 0.04 to 0.44 )\\
18&1 21 36.35&64 16 33.9&  126.199&1.596 & 31 $\pm$ 4& 10$\pm$ 1&0.91 ( 0.72 to 1.09 )\\
19&1 25  2.05&64  1  1.1&  126.602&1.385 & 24 $\pm$ 4& 13$\pm$ 3&0.48 ( 0.29 to 0.73 )\\
\hline
\hline
\end{tabular}
\end{center}
\end{table*}

\begin{table}
\begin{center}
\caption{408-1420 MHz T-T plot spectral indices
with and without Compact Sources(CS)}
\setlength{\tabcolsep}{1mm}
\begin{tabular}{cccc}
\hline
\hline
 Sp. Index   |    &$\alpha$ &  $\alpha$&  $\alpha$  \\
\hline
\hline
 Area \vline &including CS &  CS removed & manual fit\\
\hline
 a& $0.39\pm$0.33& 0.09$\pm0.70$&0.20\\
 b& $0.36\pm$0.23& 0.36$\pm0.23$&0.46\\
 c& $0.46\pm$0.09& 0.46$\pm0.09$&0.45\\
 d& 0.59$\pm$0.02& 0.59$\pm$0.02&0.59\\
 e& 0.45$\pm$0.55& 0.30$\pm$0.57&0.41\\
 f& 0.64$\pm$0.10& 0.59$\pm$0.19&0.57\\
 g& 0.34$\pm$0.14& 0.28$\pm$0.19&0.29\\
 h& 0.50$\pm$0.14& 0.29$\pm$0.24&0.19\\
\hline
All G126.2+1.6&0.42$\pm$0.09&0.30$\pm$0.08&\\
\hline
 \hline
\end{tabular}
\end{center}
\end{table}

\begin{table}
\begin{center}
\caption{Integrated flux densities and spectral indices of G126.2+1.6 and 
compact sources within the SNR}
\setlength{\tabcolsep}{1mm}
\begin{tabular}{cccc}
\hline
\hline
Freq.& G126.2+1.6& CS of G126.2+1.6 &G126.2+1.6 and CS\\
\hline
MHz &Jy&Jy &Jy\\
\hline
\hline
 408& 9.7$\pm$3.9&2.11$\pm$0.12&11.8$\pm$4.0\\
 1420&6.7$\pm$2.1&0.84$\pm$0.06& 7.5$\pm$2.2\\
\hline
$\alpha$&0.30$\pm$0.41&0.74$\pm$0.07&0.36$\pm$0.35\\
\hline
\hline
\end{tabular}
\end{center}
\end{table}

\begin{table}
\begin{center}
\caption{Integrated Flux Densities (FD) of G126.2+1.6 and of
Compact Sources within G1262.+1.6 (CSFD)}
\setlength{\tabcolsep}{1mm}
\begin{tabular}{ccccc}
\hline
\hline
Freq. &Beamwidth&FD &CSFD& references for FD\\
MHz   &arcmin   & Jy   & Jy            & \\
\hline
\hline
 83 & 59$\times$31 & 35.0 $\pm$7.0 & 7.2$\pm$0.4   &1994, Kovalenko et al.\\
 408 &3.4$\times$3.8& 9.7 $\pm$3.9* & 2.1$\pm$0.1   &this paper\\
 408 &3.5 $\times$3.9 & 12.0 $\pm$2.5* & & 1989, Joncas et al.\\
 865 & 14.5$\times$14.5 & 7.1 $\pm$ 1.6 & 1.2$\pm$0.1   & 2003, Reich et al.\\
 1420& 1$\times$1.1 & 6.7 $\pm$ 2.1* & 0.84$\pm$0.06   & this paper\\
 2695 & 4.4$\times$4.4& 4.4  $\pm$ 0.4 & &1984, F\"urst et al.\\
 2695 & 4.3$\times$4.3& 3.7  $\pm$ 0.4 & 0.52$\pm$0.06   & see text\\
 4850 & 2.6$\times$2.6& 1.7$\pm$ 0.2** & 0.34$\pm$0.07   & 1984, F\"urst et al. \\
\hline
\hline
\end{tabular} 
\\**CS have been subtracted.  ** lower limit \\
\end{center}
\end{table}

\subsection{T-T Plot Spectral Indices}
Bright compact sources affect the measured integrated flux densities for G126.2+1.6 and its 
measured spectral index. Thus we correct for the effects of compact sources. 
Table 1 lists properties of the 19 brightest compact sources which are detected
within G126.2+1.6 at both 408 MHz and 1420 MHz.   

First we discuss spectral indices between 408 MHz and 1420 MHz based on 
the T-T plot method (Turtle et al., 1962).
The principle of the T-T plot method is that spectral indices 
(T$_{\nu}$=T$_{o}$$\nu$$^{-\beta}$) are calculated from a fit of a linear 
relation to the T$_{1}$-T$_{2}$ values of all pixels within a given map region. 
T$_{1}$ is the brightness temperature of a map pixel at one frequency and 
T$_{2}$ is for the second frequency. The higher resolution image has been smoothed to the lower resolution for the T-T plot comparison. The brightness temperature spectral 
index $\beta$ is derived from the slope of the line. The error in spectral index is derived from the uncertainly in slope of the line. 
The flux density spectral index $\alpha$ 
(S$_{\nu}$$\propto$$\nu$$^{-\alpha}$) 
is related to $\beta$ by $\beta$=$\alpha$+2. Spectral index 
refers to flux density spectral index $\alpha$ in this paper unless specifically 
noted otherwise. 

For the T-T plot analysis, first a single region for the whole SNR is used, 
as shown in Fig. 1. This region yields the T-T plots shown in Fig. 2. 
Three cases are considered: using all pixels including compact sources; using all 
pixels after subtracting Gaussian fits to the compact sources 
listed in Table 1 from the images; 
and excluding compact sources. 
The compact sources are bright compared to the SNR emission. 
Since the compact sources have a steeper spectrum than the SNR, 
they are seen in the T-T plot (Fig. 2 left panel) as the steeper lines of 
points extending to higher $T_B$. Subtracting 
compact sources from the image before making
the T-T plot removes the lines of points associated with the compact sources
(Fig. 2 middle panel).
However artifacts remain in the T-T plot due to imperfect source subtraction, 
especially at 408 MHz. The next step is to completely remove regions of pixels including 
the compact sources from the analysis. Each region is taken to be a few 
beamwidths across, so that any contribution from the compact source is below 
1$\%$ of the diffuse SNR emission. Thus any artifacts associated with the compact
source are also removed. 
Generally, the last step produces the best results (rightmost plot of Fig. 2).
 Therefore, in further discussion, the second case will be not considered.

\begin{figure*}
\vspace{43mm}
\begin{picture}(60,80)
\put(-50,-50){\includegraphics{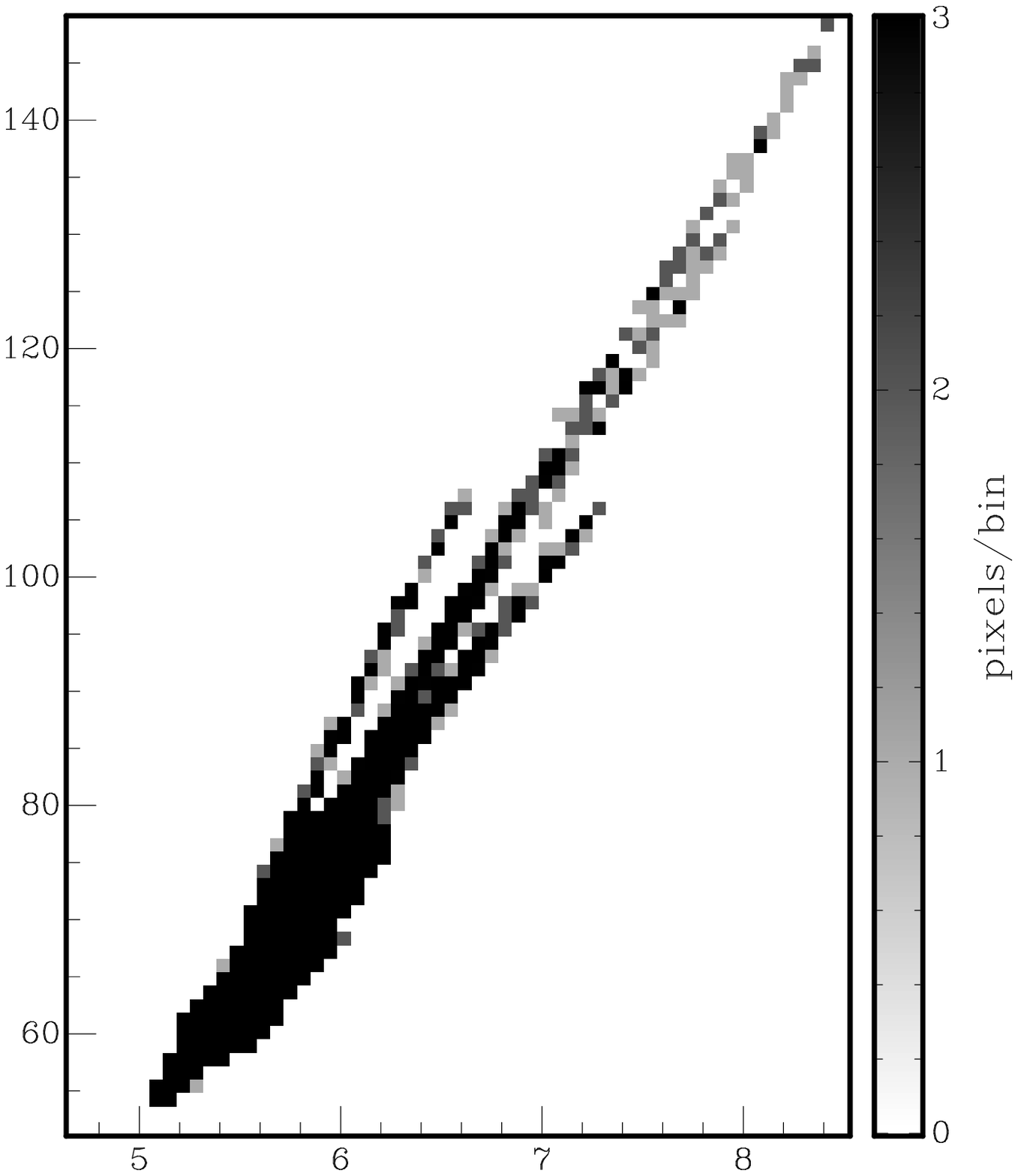}} 
\put(120,-50){\includegraphics{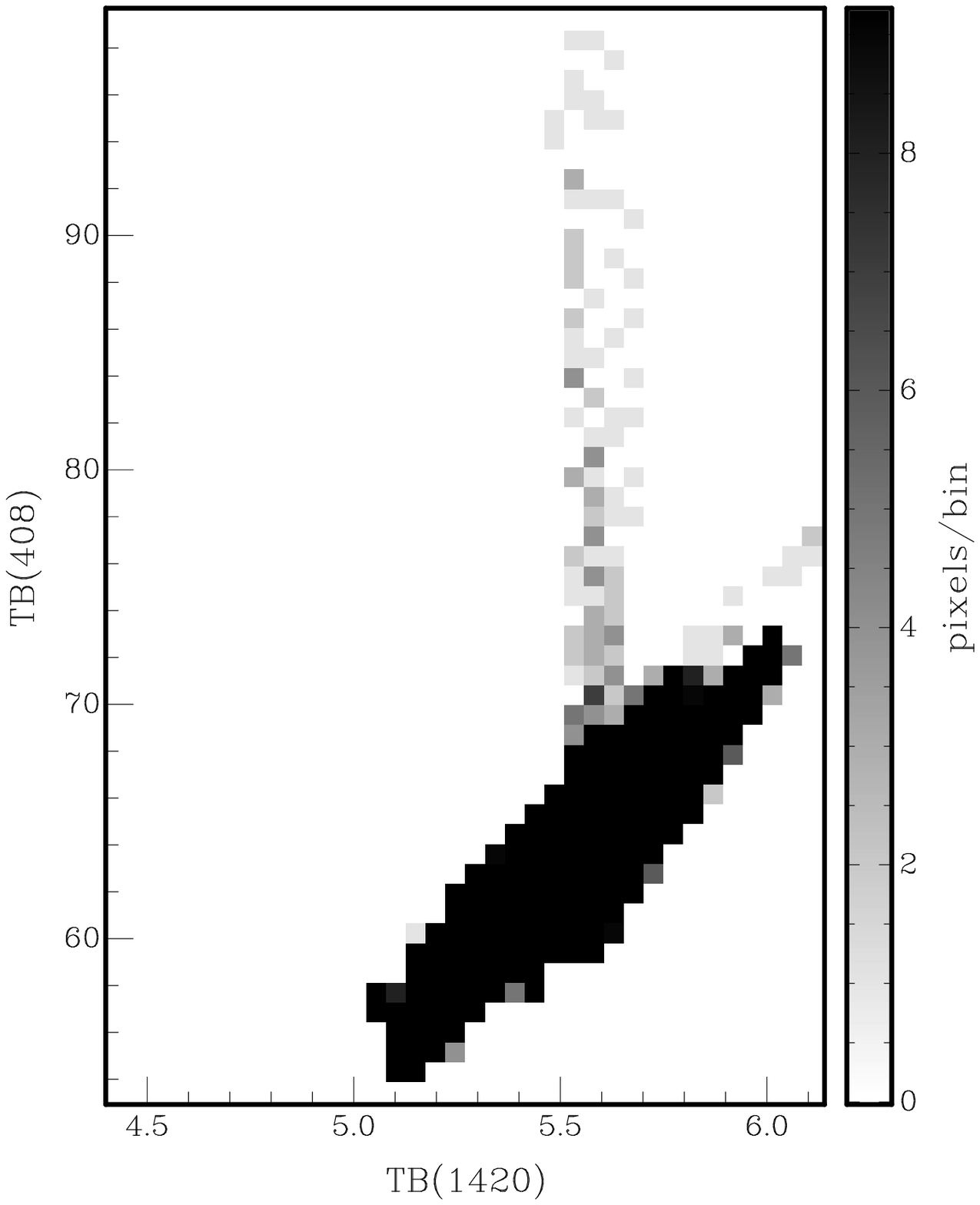}} 
\put(310,-50){\includegraphics{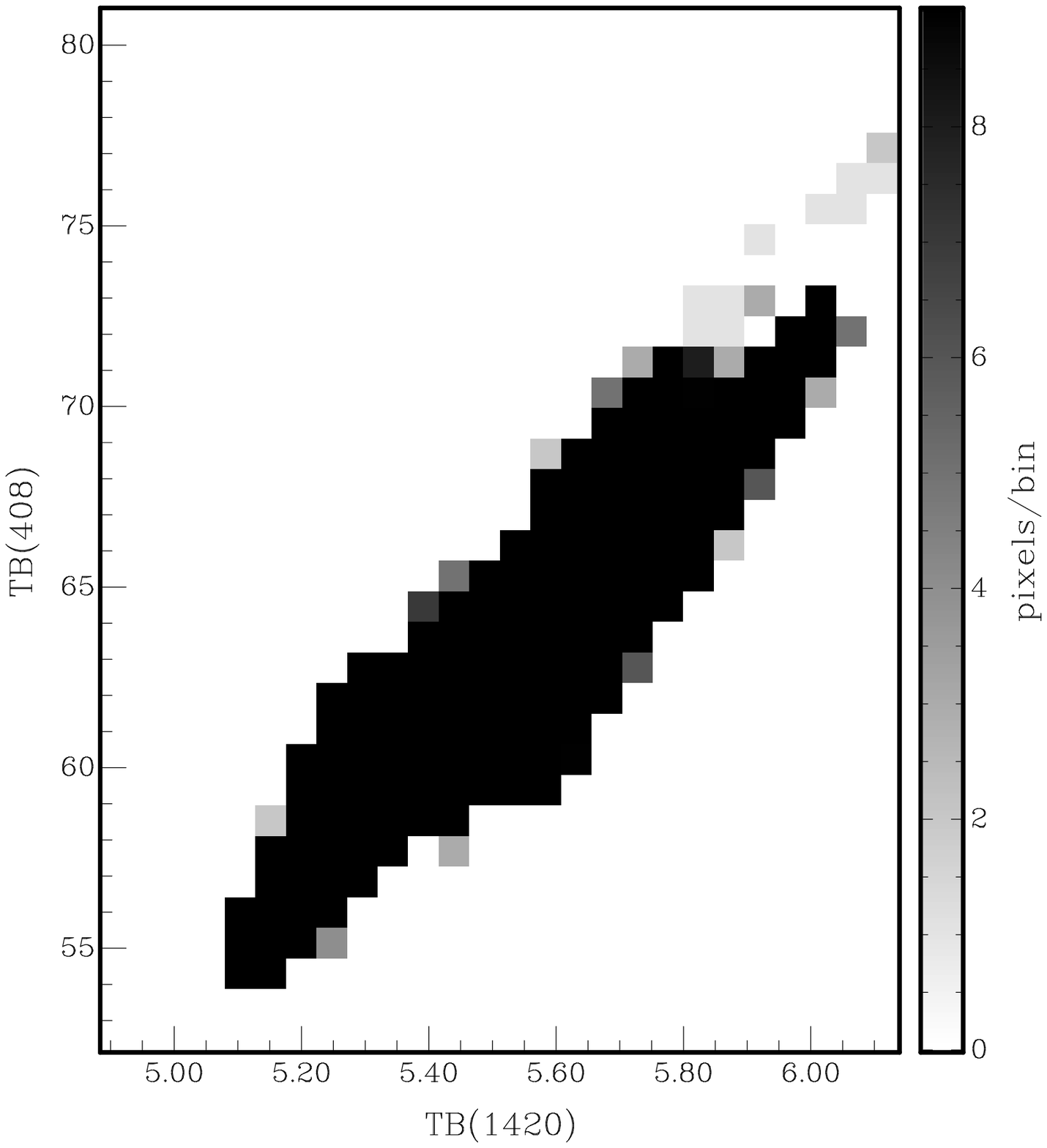}}
\end{picture}
\caption[xx]{Whole SNR 408-1420 MHz T-T plots.  From left to right: plot for map including compact sources 
($\alpha$=0.42$\pm$0.09); plot for map with Gaussian fits to compact sources subtracted 
($\alpha$=0.31$\pm$0.11); plot for compact sources removed from analysis 
($\alpha$=0.30$\pm$0.08). Coordinates units of the left plots are same as the right plot.}
\end{figure*}

Next, smaller areas (labeled a to h in the upper right panel of
Fig. 1) are selected  
to search for spatial variations in spectral index.  
Table 2 lists the results for two cases of analysis: 
including compact sources and removing compact 
sources. 
Visual inspection of the T-T plots confirms that the second method produces the most reliable results.
The compact sources' influence on the spectral index calculation 
is obvious in the T-T plots, and also seen in Table 2 for areas a, e, f, g and h. 
From now on we discuss
spectral indices derived with compact sources removed, unless specified otherwise.   
From Table 2, we see spatial variation of spectral index within G126.2+1.6: ~0.2 - ~0.6.  
 Most difference are of low significance; the most significant difference 
is between d and g (at 1.6$\sigma$).

\begin{figure}
\vspace{55mm}
\begin{picture}(0,80)
\put(-20,-20){\includegraphics{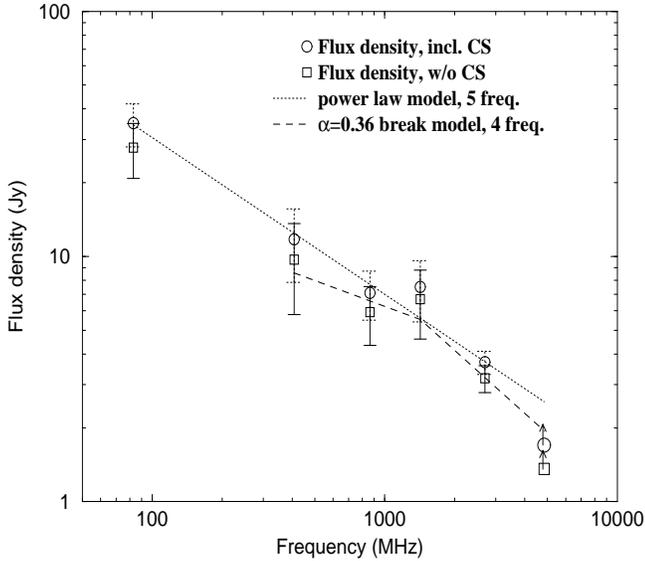}}
\end{picture}
\caption[xx]{Radio spectrum of G126.2+1.6. The value at 4850MHz is a lower limit.
The power law model has a best-fit spectral index of 0.62, $\chi^2$ = 0.99, for a fit to 5 frequencies 83-2695MHz. 
The synchrotron aging model has a best-fit spectral index of 0.36, break frequency of 1420 MHz
and $\chi^2$ =0.6, for a fit to 4 frequencies 408-2695MHz.}
\end{figure}

\begin{figure*}
\vspace{42mm}
\begin{picture}(60,60)
\put(-10,-65){\includegraphics{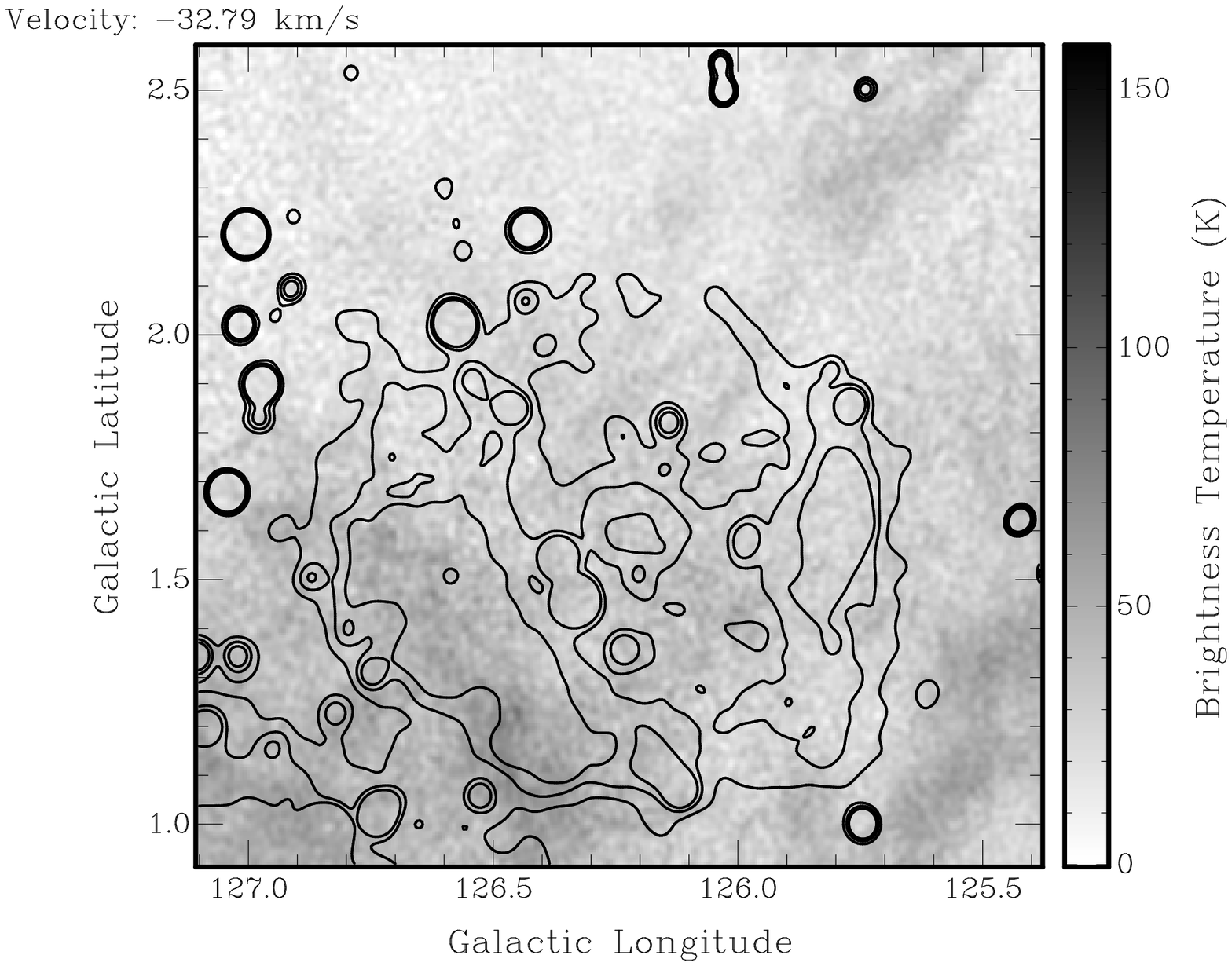}}
\put(160,-65){\includegraphics{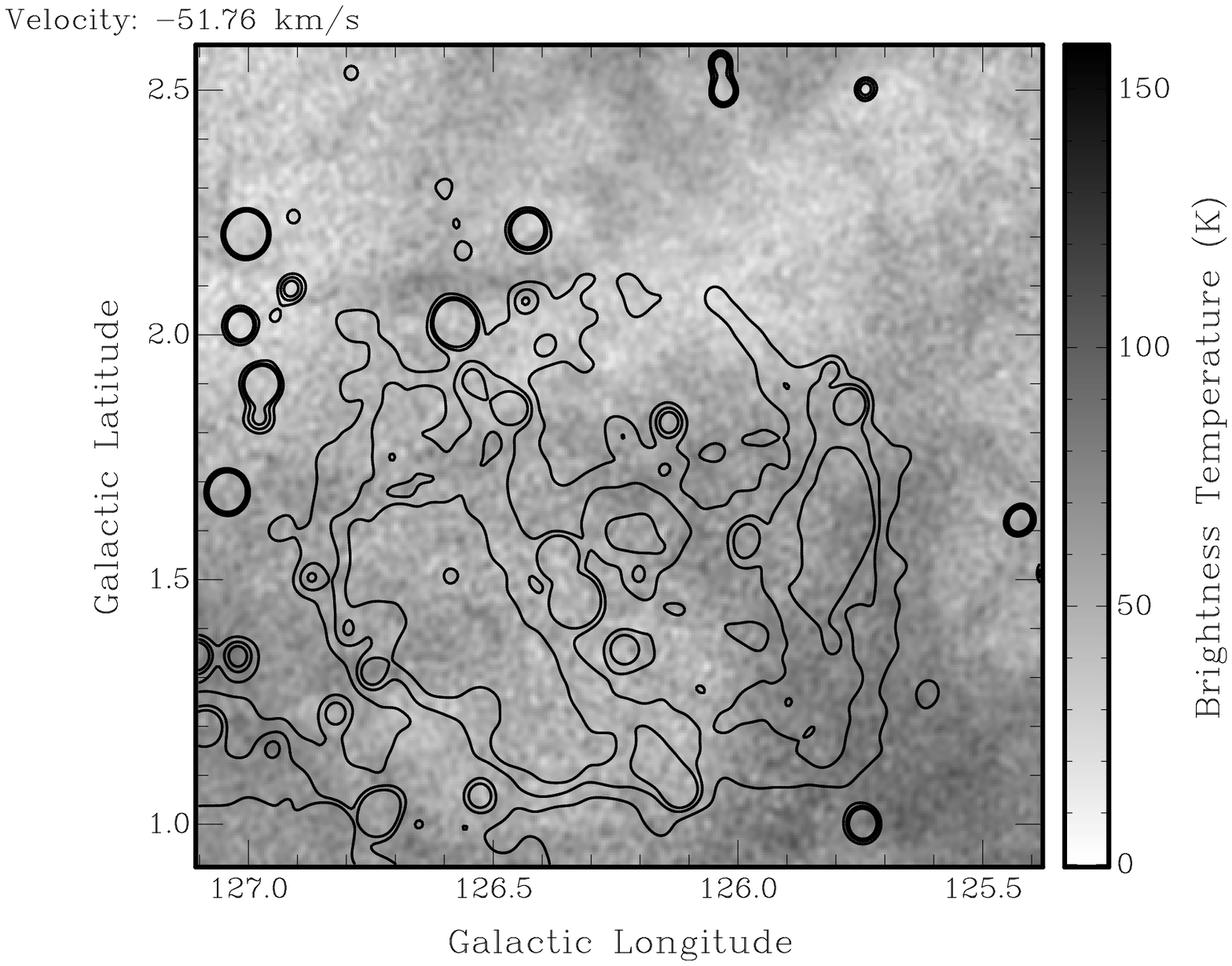}}
\put(330,-65){\includegraphics{hi111_134sm.eps}}
\end{picture}
\caption[xx]{Grey scale representation of HI emission in the field centered on G126.2+1.6. The radial velocity of the left and middle plots is indicated by the number at its top left corner. The grey scale bar of the left and middle plots has units of brightness temperature, of the right has units of 10$^{20}$ atom cm$^{-2}$. Continuum emission at 1420 MHz is indicated by contours (5.5, 5.6, 5.7,5.8,5.9 mJy beam$^{-1}$).}
\end{figure*}

\subsection{Integrated Flux Densities and Spectral Indices}

We have derived integrated flux densities of G126.2+1.6 from the 
408 MHz and 1420 MHz maps. 
Values given have diffuse background subtracted.
The resulting 408 MHz to 1420 MHz spectral index, using flux densities without compact sources, 
is 0.30$\pm$0.41.  
Table 3 lists the flux densities and spectral indices of G126.2+1.6 and 
the compact sources within G126.2+1.6. 
Compact sources contribute about 18$\%$ at 408 MHz and 13$\%$ at 1420 MHz 
to the SNR's flux densities, and have a significant effect on the spectral index. 
It is noted that the whole SNR spectral index derived from integrated flux densities is
consistent with the whole SNR spectral index (0.30$\pm$0.08) derived
by the T-T plot method.

Published integrated flux densities and errors for the SNR 
at other frequencies are given in Table 4. 
The errors are the published errors and don't include any uncertainty
due to possible differences in flux density scales. 
We have calculated total compact source flux densities for these other 
frequencies, using the 408-1420 MHz spectral index upper and lower limits and
flux densities from Table 1.
The compact sources' flux density contribution is the highest at 83 MHz at 21 $\%$. It decreases to 18$\%$ at 408 MHz, 17$\%$ at 865 MHz, 13$\%$ at 1420 MHz, 14 $\%$ at 2695 MHz and 20$\%$ at 4850 MHz. 
We have recalculated the flux density values of G126.2+1.6 at 83 MHz, 865 MHz, 2695 MHz and 
4850 MHz by subtracting the compact source flux density. 
For 408 and 1420 MHz the values in Table 4 already have compact source flux density
removed. We note that F\"urst et al (1984) gave a flux density based 
on the 2965 MHz Effelsberg image with resolution 4.4 $\times$ 4.4 arcmin, but we 
 obtain a new value from the Effelsberg 2695 MHz image  
with a little higher resolution 4.3 $\times$ 4.3 arcmin: 3.7$\pm$0.4 Jy and use this instead.
We first fit the resulting flux density values with a power-law to obtain spectral index. 
Because the flux density at 4850 MHz is a lower limit value, we do not include it in our fits. 
Figure 3 shows the corrected flux densities and the best-fit power-law.  
For the lowest 5 frequencies, there is a best fit spectral index of $\alpha$=0.62 with $\chi^2$=0.99. 
If we consider a synchrotron aging model for the G126.2+1.6 radio spectrum, the spectrum is 
steeper by 0.5 above the break frequency. The best fit to the lowest frequencies except 83 MHz has $\alpha$=0.36, $\chi^2$=0.6 and a break frequency of 1420 MHz. A detail 
analysis about the spectrum fits has  been given in discussion section.

\subsection{HI Emission}
We have searched the CGPS radial velocity range for features in the HI which might 
relate to the morphology of G126.2+1.6.
There is emission which is coincident with the boundary of G126.2+1.6 in the velocity
range -32 to -52 km/s, and only in this velocity range.
From -32 to -42 km$/$s, the HI emission is 
adjacent to the south and southeast edge of G126.2+1.6, and 
from -47 to -52 km$/$s, the HI emission is 
adjacent to the west edge. The spatial relation with the
edge of G126.2+1.6 indicates that the HI is probably associated with
the supernova remnant.
These HI observations show a good association of HI 
features with G126.2+1.6, similar to other accepted HI associations with supernova
remnants and with similar HI velocity ranges ($\sim$20km/s), e.g.  HI associated with 
the Cygnus Loop (Leahy, 2003) and with DA530 (Landecker et al., 1999). 

Fig. 4 shows maps of HI emission in two channels 
(the left and middle panels are for radial velocities -32.8 and -51.8 km$/$s 
respectively) and also integrated over channels from -32.8 to -51.8 km$/$s (the right 
panel). Each map has superimposed on it contours of continuum emission at 1420 MHz 
chosen to show G126.2+1.6. 

\section{Discussion}
\subsection{The Distance and Age to G126.2+1.6}

There are no previous reliable distance estimates for G126.2+1.6. Because of the low 
surface brightness of G126.2+1.6 and due to presence of compact sources, its flux 
density was measured with large errors. The surface brightness - diameter relations 
only gives a rough distance estimates (3.6 kpc to 7.1 kpc, Joncas et al. 1989). 
Joncas et al. (1989) discussed the distance in detail, suggested a range of 2 to 5 kpc, 
and noted that a reliable distance may be obtained based on a future measure of 
systemic radial velocity of the remnant, either from the optical lines or from 
detection of an interaction with the surrounding HI. 
The HI yields a  radial velocity of G126.2+1.6 
of -42 km$/$s. We take a galactic rotation curve with R$_{0}$=8.5 kpc, 
V$_{0}$=220 km s$^{-1}$, and V=250 km s$^{-1}$ for R $>$ 1.4 R$_{0}$, and obtain 
a distance of 5.6 kpc for G126.2+1.6.  The range of distances obtained for radial 
velocities between -39 km/s and -45 km/s is 5.3 to 5.9 kpc.

At 5.6 kpc distance, we get a mean radius of about R=64 pc for G126.2+1.6. 
From the column density of the HI associated with G126.2+1.6, we estimate an approximate density n$_0$=2$\times 10^{20}cm^{-2}$/(2R)=0.5 $cm^{-3}$. Applying a Sedov model (e.g. Cox, 1972), for a typical explosion energy of E=0.5$\times$ $10^{51}$ erg ($\epsilon_{0}$=E/$(0.75\times 10^{51}erg)$=2/3),
yields an age of 4.7$\times$10$^{5}$ yr. 
This is too old since the SNR would have completely cooled by 
that time for n$_0$=0.5 $cm^{-3}$. The shock radius when the SNR has cooled is 
$R_s^{(c)}$=24.3($\epsilon_{0}/n_{0}$)$^{5/17}n_{0}^{-2/17}$ pc, and we require 
R$\le R_s^{(c)}$. Thus we find an upper limit on n$_{0}$ is 0.1 cm$^{-3}$.  
This is consistent with the HI column density if the SNR exploded in a low density 
cavity and we are seeing the surrounding high density shell, as one sees for the 
Cygnus Loop (Leahy, 2003). The revised Sedov age for G126.2+1.6 with n$_{0}$$<$0.1 cm$^{-3}$ is less than 
2.1$\times$10$^{5}$ yr.
 We also consider a post-Sedov model. In this case (Cox, 1972), shocked gas cools and joins
a dense shell which expands slowly, $\sim$30 km/s, driven by the hot interior
of the supernova remnant. However, for n$_0$=0.5 $cm^{-3}$ and the observed size of G126.2+1.6,
the age of the supernova remnant is $\sim 10^6$ yr. This is too large, and requires
that the pre-SNR density must be much lower. So again the conclusion is that the supernova
exploded in a low density cavity in the interstellar medium, similar to the Cygnus Loop.

\subsection{Radio Spectrum}

Generally, in the spectrum plots (Fig. 3), the 1420 MHz flux density appears high and the 
83 MHz flux density appears high. 
Since our 1420 MHz map is the best so far for G126.2+1.6 the 1420 MHz one 
is the most reliable flux density. So we suspect that due to the faintness of 
the outer regions of G126.2+1.6, there is difficulty in determining the background 
level when the spatial resolution is poor. This results in increasing the uncertainty in flux densities.

The radio spectrum shown in Fig. 3 has a best-fit power-law with $\alpha$=0.62.
This is much higher than the well-determined 408-1420 MHz T-T plot spectral index and 
the flux density-based values (both about 0.30). 
Previous spectrum fits ($\alpha$=0.48, 
F\"urst et al. 1984; $\alpha$=0.58, Joncas et al. 1989; $\alpha$=0.59, 
Kovalenko et al. 1994; $\alpha$=0.77, Trushkin 2002) are not reliable due to 
inclusion of compact sources and poorer spatial resolution. For example we 
obtain $\alpha$=0.36 for 408-1420 MHz spectral index if we include compact sources 
(see Table 3). 

Here are some alternate explanations of the observed multi-frequency radio spectrum. 
One alternative is that the 83 MHz observation has too high a flux density.
If we omit the flux density at 83 MHz, we obtain a power-law fit
with $\alpha$=0.61 ($\chi^2$=0.95). This still disagrees with the
408-1420MHz spectral index.  

Another alternative is that the radio spectrum is described by a synchrotron aging 
model with resulting break in the radio spectrum.
F\"urst et al. (1984) gave a lower limit flux density at 4850 MHz and 
argued the spectrum of G126.2+1.6 is steepening above 1 GHz.   
If we include all frequencies data (except the 4850 MHz lower limit) 
we obtain $\alpha$=0.59, $\chi^2$=0.95 and break frequency 2240 MHz. 
If we use the synchrotron aging model and omit the flux density at 83 MHz, we 
obtain $\alpha$=0.36, $\chi^2$=0.56 and break frequency 1420 MHz. 
This is consistent with the 408-1420 MHz index and is a better fit than the pure power-law
model.

The last alternative we discuss, 
is that the curved spectrum is partly due to the spectral index variations within 
G126.2+1.6. 
However, when we explicitly include the flux density
from regions c and d, with their known 1420 MHz flux density and known spectral index,
we do not obtain significantly different results than above for either the power-law
or synchrotron aging models.
 
In summary, we can reconcile the 0.30$\pm$0.08 T-T plot spectral index with flux 
density-fit spectral index in two ways.
(a). Including the 83 MHz data, the radio spectrum is fit by a break spectrum with break 
frequency 1.6 MHz, and low frequency index of $\alpha$=0.4. This model is 1.3 $\sigma$ 
worse than the synchrotron aging model with free $\alpha$. 
(b) Omitting the 83 MHz data, the radio spectrum is best fit by a model with break at 
frequency 1.4 - 1.5 GHz with low frequency index $\alpha$=0.36.

\section{Conclusion}  
We present high resolution and high sensitivity images of G126.2+1.6 at 1420 MHz and 
408 MHz and show new HI-line emission data of G126.2+1.6 in this paper. The 1420 MHz 
continuum image is the best image of G126.2+1.6 at any frequency yet.  Flux densities 
at 408 MHz and 1420 MHz are obtained.  The T-T plot spectral index is 0.30$\pm$0.08 and 
the flux density-based spectral index is 0.30$\pm$0.41 between 408 MHz and 1420 MHz. By 
our analysis, a multi-frequency spectral index of about 0.36 is obtained using a 
synchrotron aging model. 
There is evidence at 1.6$\sigma$ for spatial variations in spectral index
in G126.2+1.6 between about 0.3 and 0.6. 
The association between HI brightness features and the SNR's structure suggests
a distance of 5.6 kpc for G126.2+1.6.  The estimated Sedov age for G126.2+1.6 is less than 2.1$\times$$10^{5}$ yr.
  
\begin{acknowledgements}
We acknowledge support from the Natural Sciences and Engineering Research Council of Canada. 
W.W. Tian thanks the National Natural Science Foundation of China for support.  
The DRAO is operated as a national facility by the National Research Council of Canada.  
The Canadian Galactic Plane Survey is a Canadian project with international partners. 
\end{acknowledgements}

\end{document}